\documentclass[preprint,12pt]{elsarticle}

\usepackage{amssymb}
\usepackage{float}
\usepackage{adjustbox}
\usepackage{amsmath}
\usepackage{tabularx} 

\journal{arxiv}

\begin{document}

\begin{frontmatter}



\title{Integrating Dynamic Correlation Shifts and Weighted Benchmarking in Extreme Value Analysis}


\author[inst1]{Dimitrios P. Panagoulias} \ead{panagoulias_d@unipi.gr}
\author[inst2]{Elissaios Sarmas}\ead{esarmas@epu.ntua.gr}
\author[inst2]{Vangelis Marinakis}\ead{vmarinakis@epu.ntua.gr}
\author[inst1]{Maria Virvou}\ead{mvirvou@unipi.gr}
\author[inst1]{George A. Tsihrintzis}\ead{geoatsi@unipi.gr}
\affiliation[inst1]{organization={Department of Informatics, University of Piraeus},
            addressline= {Karaoli ke Dimitriou 80}, 
            city={Piraeus},
            postcode={18534}, 
            country={Greece}}

\affiliation[inst2]{organization={Decision Support Systems Lab, School of Electrical and Computer Engineering National Technical University of Athens},
            addressline={ Ir. Politechniou 9, Zografou }, 
            city={Athens},
            postcode={157 73}, 
            country={Greece}}

\tnotetext[t1]{This work has been submitted to the IEEE for possible publication. Copyright may be transferred without notice, after which this version may no longer be accessible.}

\begin{abstract}
This paper presents an innovative approach to Extreme Value Analysis (EVA) by introducing the Extreme Value Dynamic Benchmarking Method (EVDBM). EVDBM integrates extreme value theory to detect extreme events and is coupled with the novel Dynamic Identification of Significant Correlation (DISC)-Thresholding algorithm, which enhances the analysis of key variables under extreme conditions. By integrating return values predicted through EVA into the benchmarking scores, we are able to transform these scores to reflect anticipated conditions more accurately. This provides a more precise picture of how each case is projected to unfold under extreme conditions. As a result, the adjusted scores offer a forward-looking perspective, highlighting potential vulnerabilities and resilience factors for each case in a way that static historical data alone cannot capture.
By incorporating both historical and probabilistic elements, the EVDBM algorithm provides a comprehensive benchmarking framework that is adaptable to a range of scenarios and contexts.
The methodology is applied to real PV data, revealing critical low - production scenarios and significant correlations between variables, which aid in risk management, infrastructure design, and long-term planning, while also allowing for the comparison of different production plants. The flexibility of EVDBM suggests its potential for broader applications in other sectors where decision-making sensitivity is crucial, offering valuable insights to improve outcomes.

\end{abstract}




\end{frontmatter}


\section{Introduction}

Understanding and managing extreme events is crucial across multiple domains, as these rare occurrences often carry significant risks and consequences. In finance for example, extreme events such as market crashes or rapid shifts in asset prices can lead to substantial financial losses, making the prediction and mitigation of such extremes critical for portfolio management and risk assessment. Consequently, extreme value theory (EVT) has been widely applied to model financial market tail risks (e.g., \cite{novak2011extreme,longin2016extreme,nolde2021extreme}), helping institutions prepare for catastrophic losses by better managing their capital reserves.

In healthcare, extreme events like rare medical conditions, pandemics, or extreme fluctuations in patient health metrics (e.g. in blood glucose levels) require swift responses \cite{chiu2018mortality,panagoulias2022extremec, panagoulias2022extremeb}. The ability to predict and mitigate these extremes improves patient care and health outcomes. Consequently, EVA has been applied to predict rare but severe health crises, like extreme blood pressure spikes in critically ill patients, which allows for more informed treatment strategies.

In the energy sector, particularly with regard to renewable energy, understanding extreme events is equally important. Fluctuations in energy production due to extreme weather conditions (e.g., prolonged cloudy periods or intense storms) can significantly affect the reliability of energy grids \cite{ghorbel2014energy,xie2014extreme,doherty2011extreme}. Renewable sources, such as solar and wind, are inherently variable, and extreme lows in production can lead to energy shortages. Therefore, applying EVA in this context enables energy providers to better predict, plan for, and mitigate risks associated with extreme energy production events.

This paper introduces a novel methodology, the Extreme Value Dynamic Benchmarking Method (EVDBM), which enhances the application of Extreme Value Analysis (EVA) to detect and analyze rare, high-impact events across various fields. EVDBM integrates extreme value theory with the innovative Dynamic Identification of Significant Correlation (DISC)-Thresholding algorithm, allowing for a deeper examination of how key variables behave under extreme conditions. This combination enables decision-makers in various fields, including the aforementioned ones, to better understand the correlations between critical factors during extreme scenarios, whether they involve market crashes, medical emergencies, or energy production shortfalls and project how the related conditions while adjust in the future.

The EVDBM Algorithm provides a robust quantitative mechanism for comparing different use cases by generating a final benchmarking score based on weighted performance during extreme events. This approach  considers the frequency of past extreme events and incorporates the predicted severity and likelihood of future occurrences to project how related conditions to the EVA dependent variable behave. By integrating both historical data and probabilistic projections, the algorithm enables a more meaningful comparison of cases under projected stress conditions, offering a forward-looking evaluation of performance. This comprehensive framework is adaptable to a wide range of scenarios and contexts, making it particularly valuable for applications where understanding resilience to extreme conditions is critical.

This scoring system offers a valuable tool for assessing and comparing similar situations, such as different financial portfolios during market downturns, patient responses during health crises, or energy systems under adverse weather conditions. By extracting and quantifying correlations between variables, this methodology provides insights into the overall changes that drive extreme conditions, helping sectors make informed decisions on risk management, planning, and resource allocation.

\section{Related Work}
In \cite{Arsenault}, the authors used extreme value theory for the estimation of risk in finite-time systems, especially for cases when data collection is either expensive and/or impossible. For the monitoring of rare and damaging consequences of high blood glucose, EVA has been deployed using the block maxima approach~\cite{Szigeti}. More examples of application of EVT can be found in the recent literature, and here we only report those considered more relevant to our~research. 

Extreme value analysis in energy production and consumption, particularly in the context of renewable sources like solar and wind power, is essential for balancing energy demands. Various studies \citep{chen2020modeling,westerlund2019extreme} have focused on modeling the production of solar and wind power using the Peaks over Threshold (POT) method. These method approximate the frequency of peaks above a certain threshold and determine the distribution of these peaks' sizes. Different clustering methods for analysis to determine optimal fit. Time and longer data periods are usually necessary to account for seasonal effects fully. This type of analysis is crucial for understanding and managing the variability and unpredictability inherent in renewable energy sources.

In \cite{ahmed2024extreme}, estimators for the extreme value index and extreme quantiles in a semi-supervised setting were developed, leveraging tail dependence between a target variable and co-variates, with applications to rainfall data in France. In \cite{gilleland2013software}, the authors review available software for statistical modeling of extreme events related to climate change. In \cite{makkonen2019improved}, a novel method is proposed for estimating the probability of extreme events from independent observations, with improved accuracy by minimizing the variance of order-ranked observations, eliminating the need for subjective user decisions.

Extreme value analysis (EVA) has been used in partial coverage inspection (PCI) to estimate the largest expected defect in incomplete datasets, though uncertainties in return level estimations are often under-reported \cite{benstock2017extreme}.
\subsection{Theory  of Extreme Value~Analysis}
\label{sec:EVA}
In this section, the~key characteristics of extreme value theory are highlighted. Extreme value analysis (EVA) can be approached from two different angles. The~first one refers to the block maxima (minima) series. According to block maxima (minima), the~annual maximum (minimum) of time series data is extracted, generating an annual maxima or minima series, simply referred as AMS. The~analysis of the AMS datasets are most frequently based on the results of the Fisher--Tippett--Gnedenko theorem, which leads to the fitting of the generalized extreme value distribution. A~wide range of distributions can also be applied. The~limiting of distributions for the maximum (minimum) of a collection of random variables from the same distribution is the basis of the examined theorem~\cite{coles2001introduction}.

The peak-over-threshold (POT) methodology is the second approach used in EVA In~POT, a~sorted series is analyzed, first identifying the peak values that exceed a given threshold in a given set of records. The~analysis usually involves the fitting of two distributions. One concerns the number of events covering the time period or space analyzed, and the other concerns the selected size of extracted peaks.  As~per the Pickands--Balkema--De Haan theorem, the~POT extreme values asymptotically follow the generalized Pareto distribution family, and a Poisson distribution is used for the total number of events~\cite{xu2014proceedings}. The~return level (R.V.) of the extreme values can be approximated from the fitted distribution. The~value expected or return value is equal to or exceeds the threshold on average once every interval $T$ of time or space with a probability of $1/T$.

PDF refers to the probability density function of the continuous random variable, which, at~any given point in the examined space, can provide the relative likelihood that the random variable is located near the sample space~\cite{coles2001introduction}.

Estimators based on asymptotic extreme value theory have been proposed, and their performances were theoretically evaluated and verified via Monte Carlo simulation as faster alternatives for estimation of the parameters of alpha-stable impulsive interference in~\cite{tsihrintzis1996fast}.

The shape of the probability distribution is calculated via the $L$-moments. The~$L$-moments represent linear combinations of order statistics ($L$-statistics) similar to conventional moments. They are used to calculate quantities analogous to standard deviation, skewness and kurtosis, and can thus be termed $L$-scale, $L$-skewness and $L$-kurtosis. Therefore, they summarize the shape of the probability distribution:
 
    
\begin{math} L_4= \frac{n * \Sigma n_i(Y_i - \tilde{Y})^4 }{(\Sigma n_i(Y_i –\tilde{Y})^2)^2} \end{math}.

\begin{math} L_4\end{math}= $L$-kurtosis.

\begin{math}Y_i\end{math}: $i$th Variable of the~distribution.

\begin{math}\tilde{Y}\end{math}: Mean of the~distribution.

$n$: Number of Variables in the~distribution.

\begin{math}\tilde{\mu}_{3}=\frac{\Sigma^N_i(X_i-\tilde{X})^3}{(N-1)*\sigma^3}\end{math}.
 
\begin{math}\tilde{\mu}_{3}\end{math}=$L$-skewness .

\begin{math}N \end{math}= number of variables in the~distribution.

\begin{math}X_i\end{math} = random~variables.

\begin{math}\tilde{X} \end{math}=mean of the~distribution.

\begin{math}\sigma \end{math} = standard deviation.
namely.
\subsection{Pearson correlation}
The Pearson correlation coefficient \cite{jebli2021prediction,benesty2008importance,sedgwick2012pearson}, often denoted as r, is a measure of the linear relationship between two variables. The Pearson correlation coefficient quantifies the degree to which two variables $X$ and $Y$ are linearly related. It ranges from -1 to 1, where:

    $r$ = 1 indicates a perfect positive linear relationship (as X increases, Y increases proportionally)
    
     $r$ = -1 indicates a perfect negative linear relationship (as X increases, Y decreases proportionally)
    
  r = 0 indicates no linear relationship between X and Y

The Pearson correlation coefficient between two variables X and Y is calculated as:
$r = \frac{\text{Cov}(X, Y)}{\sigma_X \sigma_Y}$
where: 

    $\text{Cov}(X, Y)$ is the covariance between $X$ and $Y$,
    
    $\sigma_X$ and $\sigma_Y$ are the standard deviations of X and Y, respectively.

The covariance measures how two variables move together and is defined as:
$
\text{Cov}(X, Y) = \frac{1}{n} \sum_{i=1}^n \left(X_i - \mu_X\right)\left(Y_i - \mu_Y\right)
$
where: 

	n is the number of data points,
 
$X_i$ and $Y_i$ are the individual values of the variables $X$ and $Y$,

 $\mu_X$ and $\mu_Y$ are the means (averages) of $X$ and $Y$, respectively.

 Lastly, the standard deviation of a variable $X$ is calculated as the measure of how spread out the values of $X$ are and is given by: 

 $
\sigma_X = \sqrt{\frac{1}{n} \sum_{i=1}^n \left(X_i - \mu_X\right)^2}
$

    
\begin{math} L_4= \frac{n * \Sigma n_i(Y_i - \tilde{Y})^4 }{(\Sigma n_i(Y_i –\tilde{Y})^2)^2} \end{math}.

\begin{math} L_4\end{math}= $L$-kurtosis.

\begin{math}Y_i\end{math}: $i$th Variable of the~distribution.

\begin{math}\tilde{Y}\end{math}: Mean of the~distribution.

$n$: Number of Variables in the~distribution.

\begin{math}\tilde{\mu}_{3}=\frac{\Sigma^N_i(X_i-\tilde{X})^3}{(N-1)*\sigma^3}\end{math}.
 
\begin{math}\tilde{\mu}_{3}\end{math}=$L$-skewness .

\begin{math}N \end{math}= number of variables in the~distribution.

\begin{math}X_i\end{math} = random~variables.

\begin{math}\tilde{X} \end{math}=mean of the~distribution.

\begin{math}\sigma \end{math} = standard deviation.


\subsection{Normalization}
Normalization refers to the process of scaling variables so that they fit within a common range (e.g., [0, 1] or mean 0 and standard deviation 1). In many applications, normalization preserves the relative differences between variables but still retains their units in some form (although scaled). It is often used in data science, statistics, and machine learning, where the goal is to make variables comparable by bringing them onto the same scale \cite{carrino2016data,vafaei2018data,hancock1988data,abdi2010normalizing}. Common approaches in normalization aimed at transforming the variables into a comparable, dimensionless format are the following
\begin{itemize}
    \item Min-Max Normalization (Feature Scaling): Min-max normalization is a rescaling technique where variables are linearly scaled to a specific range, often [0, 1]. The formula is:  $x_{\text{normalized}} = \frac{x - \min(X)}{\max(X) - \min(X)}$. 
    Where:
    \begin{itemize}
        \item  $x$  is the original value of the variable.
        \item  $\min(X)$  and  $\max(X)$  are the minimum and maximum values of the variable  $X$ .
    \end{itemize}
    \item Z-Score Normalization (Standardization): Z-score normalization transforms each variable by subtracting the mean and dividing by the standard deviation. The formula is: $x_{\text{standardized}} = \frac{x - \mu_X}{\sigma_X}$
    Where: 
    \begin{itemize}
        \item 	 $x$  is the original value
        \item  $\mu_X$ is the mean of the variable  $X$ 
        \item  $\sigma_X$  is the standard deviation of  $X$ 
    \end{itemize}

\end{itemize}

\section{Materials and Methods}
In this section we describe the different methods used and how they are adapted into new a novel methodology.

As a basis to our approach we utilize Extreme Value Analysis (EVA) to identify and understand rare event scenarios, thereby aiding in informed decision-making and optimizing management strategies. Evaluating the circumstances \cite{panagoulias2023circumstance} related to extreme values and looking for patterns to inform on decisions and consists of an elaborated process and two algorithmic approaches aimed at identifying, explaining and expanding on extreme circumstances.

\subsection{Process}

\begin{figure}[H]
    \centering
    \includegraphics[width=1.1\linewidth]{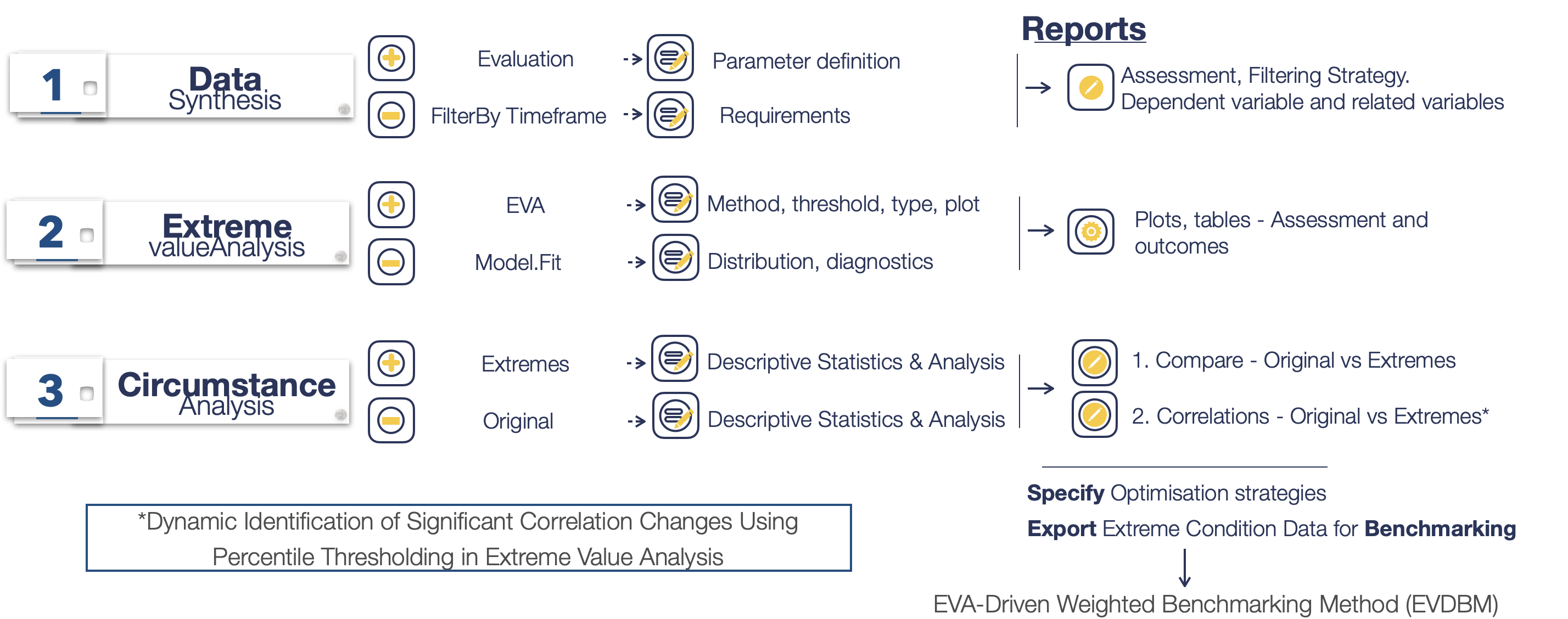}
    \caption{Extreme Value Dynamic Benchmarking Method (EVDBM)}
    \label{fig:eva}
\end{figure}

In figure \ref{fig:eva} we outline our proposed methodology named Extreme Value Dynamic Benchmarking Method (EVDBM) where the dynamic identification of significant correlation (DISC) changes is a component. It is a three (3) step process of analytical milestones, presented as a process. In more detail:

\begin{enumerate}
    \item Data Synthesis
    \begin{itemize}
        \item Evaluation: This step involves assessing the available data to define key parameters that will be used in the analysis. This might include identifying which data points are relevant to the study and determining the scope and nature of the data to be synthesized.
        \item Filter by timeframe: This process filters the data according to specific time periods. The requirements for the timeframe are established, which could involve selecting data from particular years, months, or days that are relevant to the analysis objectives.
        \begin{itemize}
        \item \textbf{Report}: Assessment, Filtering Strategy, Dependent Variable and Related Variables: This report synthesizes the findings from the data synthesis and extreme value analysis, detailing the assessment of the data, the strategy used to filter it, and an analysis of the dependent variable in relation to other relevant variables.
        \end{itemize}
    \end{itemize}
    \item Extreme Value Analysis
    \begin{itemize}
        \item EVA (Extreme Value Analysis): In this stage, the method for conducting Extreme Value Analysis is selected, including the determination of thresholds, types of extremes (such as maximum or minimum values), and how the results will be visualized (e.g., through plots).
        \item Model Fit: This step involves fitting a statistical model to the data, focusing on the distribution of extreme values. Diagnostic checks are likely performed to ensure the model fits the data appropriately and to validate the assumptions of the statistical analysis.
        \begin{itemize}
        \item \textbf{Report}: Plots, Tables - Assessment and Outcomes: Visual and tabular representations of the data are provided, which support the assessment and help communicate the outcomes of the analysis.
        \end{itemize}
    \end{itemize}
    \item Circumstance Analysis
    \begin{itemize}
        \item Extremes: Analysis of the extreme values using descriptive statistics and other analytical methods. This part of the process aims to understand the behavior of the data at its extremes.
        \item Original: Similar analytical techniques are applied to the original dataset (not just the extremes) to provide a comprehensive understanding of the overall characteristics of the examined data.
        \begin{itemize}
            \item \textbf{Report}: Compare - Original vs Extremes and Correlations - Original vs Extremes: This involves a comparative analysis between the (a) datapoints in the normal scenario  and the extreme datapoints, as well as an (b) examination of the correlations between them to identify patterns or relationships. 
            For (a) we assume that $T$  represents any descriptive statistic to be computed (e.g., mean, median, variance, standard deviation). \( x_1, x_2, \dots, x_n \) represent the values under the extreme scenario and \( y_1, y_2, \dots, y_n \) represent the values under the normal scenario.
            Then: $\Delta T = T(x) - T(y)$  such that mean can be represented as 
        $\Delta \bar{v} = \bar{v}{\text{extreme}} - \bar{v}{\text{normal}}$ and so on.

            For the streamlining of (b) we have constructed a novel algorithm as Dynamic Identification of Significant Correlation Changes Using Percentile Thresholding in Extreme Value Analysis explained in the following section.
        \end{itemize}
    \end{itemize}
\end{enumerate}

Lastly, in the ``Specify Optimization Strategies," the outcome of the reports should inform on strategies to optimize processes, systems, or models based on the insights gained from the analysis. To further expand on this, we have constructed two novel algorithms that expand on EVA, by (1) dynamically identifying significant correlation changes which attached to the circumstance analysis (step 3) and (2) by using those extremes as a benchmarking basis for different use cases. 

\subsection{Dynamic Identification of Significant Correlation Changes Using Percentile Thresholding in Extreme Value Analysis (DISC-Thresholding)}
Below is the generalized mathematical formula that incorporates a dynamic threshold for identifying significant correlation changes, based on a High Positive Correlation (HPC) and a High Negative Correlation (HNN).

Let:

 $\Delta \rho_{ij}$ be the change in correlation between variables  $X_i$  and  $X_j$, defined as:

$\Delta \rho_{ij} = \rho_{ij}^{\text{extreme}} - \rho_{ij}$ (1)

where:

\begin{itemize}
    \item  $\rho_{ij}^{\text{extreme}}$ is the Pearson correlation coefficient between  $X_i$  and  $X_j$  in the extreme dataset 
    \item $ \rho_{ij} $ is the Pearson correlation coefficient between  $X_i$  and  $X_j$  in the general dataset
\end{itemize}

We define thresholds based on the 90th (or other) and 10th (or other) percentiles of the distribution of  $\Delta \rho_{ij}$ :
\begin{itemize}
    \item  $P_{90}(\Delta \rho)$ The 90th percentile of all $\Delta \rho_{ij}$  values, representing high positive changes in correlation (2)
    \item  $ P_{10}(\Delta \rho)$ The 10th percentile of all  $\Delta \rho_{ij}$  values, representing high negative changes in correlation (3)
\end{itemize}
Then, we define the classification of significant changes in correlation as follows:

$\text{Classified}(\Delta \rho_{ij}) =
\begin{cases}
\text{HPC (High Positive Correlation)}, & \text{if } \Delta \rho_{ij} (1) > P_{90}(\Delta \rho) (2)\\ 
\text{HNC (High Negative Correlation)}, & \text{if } \Delta \rho_{ij} (1)< P_{10}(\Delta \rho) (3)\\
0, & \text{otherwise (Not Significant as 0)}
\end{cases}$
In detail: 
 \begin{itemize}
     \item 	$\Delta \rho_{ij}$ (1) represents the change in the pairwise correlation between variables  $X_i$  and  $X_j$  when comparing extreme events to normal events.
    \item HPC (High Positive Correlation) occurs when the change  $\Delta \rho_{ij}$  exceeds the 90th percentile, indicating a significant increase in correlation during extreme events.
    \item	HNC (High Negative Correlation) occurs when the change  $\Delta \rho_{ij}$ is below the 10th percentile, indicating a significant decrease in correlation during extreme events.
	\item Values between the 10th and 90th percentiles are considered not significant.
 \end{itemize}

Thus, the entire matrix of correlation differences is classified as HPC, HNC, or Not Significant depending on the relative magnitude of  $\Delta \rho_{ij}$  compared to the dynamically computed thresholds  $P_{90}$  and  $P_{10}$.

\subsection{EVA-Driven Weighted Benchmarking Algorithm for Performance}
This process is initialized by first describing related conditions  under extreme conditions and non-extreme circumstances, allowing for benchmarking and guiding operational or strategic decisions.

$V$  represents a set of key variables, related to the dependent variable examined using EVA.

\begin{math}V = \{V_1, V_2, \dots, V_n\}\end{math} (4)

 $C_{\text{extreme}}(V_i)$  and  $C_{\text{normal}}(V_i)$  are the counts or statistics (e.g., average or sum) of variable  $V_i$  under extreme low (or high) conditions and normal conditions, respectively, for a given use case.
$C_{\text{extreme}}^1(V_i)$  and  $C_{\text{extreme}}^2(V_i)$ are the values for two different use cases, let say Case 1 and Case 2, under extreme (low or high) conditions. Moving forward, the \textbf{benchmarking algorithm process} ensues. 

\newpage
\textbf{Benchmarking algorithm process:}

\begin{itemize}
    \item Step 1: Scaling factor, that incorporates historical occurrences and accounts for future risk and adjusting for Stationarity using the projected return values

For the \textbf{historical occurrences} we can compute a scaling factor  S\_j  for each case (c) that adjusts its score based on the frequency and intensity of extreme events by normalizing the number of extreme events:

$E_c = \frac{N}{\sum_{k=1}^{m} N_k}$ (5)

Where:

\begin{itemize}
    \item N  is the number of extreme events for a case (c) 
    \item $\sum_{k=1}^{m} N_k$  is the total number of extreme events across all cases
\end{itemize}

The EVA distributions provides a return level or exceedance probability, which informs on the likelihood that a given threshold will be exceeded in a specific time period. 
The probability of exceeding a threshold  $P(X > x\_0)$ accounts for the future risk and can be estimated as follows:

$P_j(X>x) = 1/T_{\text{return}}\times x(T)^{CI} $  (6)

Where: 
\begin{itemize}
    \item $T\_{\text{return}}$  is the return period, which is the average time between extreme events that exceed the threshold  $x$ .
    \begin{itemize}
        \item Such that if $T\_{\text{return}}$ =5 (years), then the probability of an extreme event happening in any given year is $1/5=0.2$
    \end{itemize}
    \item and \begin{math}
    x
\end{math} is the Return Value per \begin{math}
    T
\end{math} depending on the associated \textbf{confidence intervals} - \begin{math}
    CI[lower, upper]
\end{math}. Using the projected return values add a more dynamic approach to the stationary (normalized) related circumstances
\end{itemize}
To account both for historical occurrences and the associated probabilities the final scaling factor is:
\begin{math}S_j = E_c \times P(X>x)\end{math} (7)

Thus for the final calculation we also consider the upper and lower limits and as such the final scaling factor is formulated in a way to include all scenarios, thus:
\begin{itemize}
    \item Return Value: 
    $P_j(X>x_0)$  (a)
    \item Lower CI: $P(X>x_\text{lower})$ (b)
    \item Upper CI: $P(X>x_\text{upper})$ (c)
\end{itemize}

\item Step 2: Weighting Factor and Normalization if required

If variables used are of different measuring unit, normalization is applied. To account for the fact that different variables may have varying degrees of influence, we introduce weights for each variable. The weights reflect the importance of each  variable under extreme conditions so that: 

$b(V_i) =  w_i \times{C_\text{extreme}}(V_i)$ (7)

Where:
$w_i$  is the weight associated with the \textbf{mean} of the variable  $V_i$ and   $V_i$. The weights allow for prioritization on the more influential variables in the benchmarking process.
\item Step 3: \textbf{Benchmarking Score (B)}

The Benchmarking Score enumerates the final score per examined use case and the highest (or lowest score) suggests that there are better conditions associated to that case. We calculate the benchmarking score per projected year(s) as reported by the EVA, using the associated scaling factor. Calculations are considered for return value and the associated confidence intervals. This allows for visual representation of process to inform on more accurate judgments and comparisons. Thus considering (6), (7) and (8): 
$B_{\text{i}} = S_j\times\sum_{i=1}^{n} b(V_i)$ (9)
for (a), (b) and (c).

\item Step 4: \textbf{Visualization of Benchmarking Scores as a time series}

For the visualization of the Step 3 we use a logarithmic scale for the x-axis representing return periods to improve the clarity and interpretability of the benchmarking plots. Given that the return periods span several orders of magnitude (e.g., 1 to 1000 years), a logarithmic scale allows for a more even distribution of points across the axis, making it easier to observe trends over time. This approach ensures that smaller periods (e.g., 1, 2, 5 years) and larger periods (e.g., 100, 500, 1000 years) are both visible in the same plot without compressing the data, which would occur on a linear scale.

\end{itemize}

\section{Application of methods}
In this section we apply the EVDBM methodology on Photovoltaic (PV) data taken from two different PV Plants \cite{sarmas2023short, sarmas2022transfer}. We follow the three step process outlined, without strictly proposing an optimization strategy but rather outlining a few as an example of applicability. We ill also first introduce some ``dummy" weights in order to conclude with a final benchmarking score. 

In our analysis we considered total production in kwh of the PV production farms. We considered as time range of interest the peak hours which occur during midday, thus a time range of 3 hours from 13:00 to 16:00 was examined. The data were taken from \cite{sarmas2024photovoltaic, ilias2023unsupervised} and are take from 2 PV plants situated in various regions of Portugal as provided from the non-profit organization, Coopérnico. More descriptive statistics per plant examined to be provided in the following sections.  We analyze and apply EVA on the production below the 25th percentile. The variables related to the production to be analyzed are shown in table \ref{tab:data_columns}, for context the Produzida-Production row, which refers to the Dependent variable for the EVA, is provided. We also attach the main analytical manually pre-set constants for reference, as Time-Range and Percentile, to be applied in both use Cases. 

\begin{table}[H]
\centering
\begin{tabular}{cll}
\hline
\textbf{\#} & \textbf{Variables} & \textbf{Non-Null Count \& Dtype} \\
\hline
1 & \textbf{Humidity} v1 &  non-null float64 \\
2 & \textbf{Temperature} v2 &  non-null float64 \\
3 & \textbf{cloudcover} v3 &  non-null float64 \\
4 & \textbf{windspeedKmph} v4 &  non-null float64 \\
5 & \textbf{Solar} w/m\textsuperscript{2} v5&  non-null float64 \\
6 & \textbf{Diffuse Solar} w/m\textsuperscript{2} v6&  non-null float64 \\
7 & \textit{Produzida-Production} v7 &  non-null float64 \\
\hline \hline
\textbf{\#} & \textbf{Constant} & \textbf{Value} \\
\hline
- & Time-Range & 13:00-14:00 \\
- & Percentile & 25\%\\
-& EVA-method &POT (Peaks Over Threshold) \\ 
-& Extremes type & Low \\ 
-& Negative Significance & $<10\%$ \\
- &  Positive Significance & $>90\%$ \\
\hline
\end{tabular}
\caption{Data Columns and Pre-set Constants Overview}
\label{tab:data_columns}
\end{table}

. 
\subsection{EVDBM application Use Case 1: ``Zarco" Data}

\subsubsection{Data Synthesis}
\textbf{Report:} From a total of 21932 data, 3656 were related to peak time production 
During the peak hours the ours the data to be analyzed are 3656 with a mean production of 23.5 kwh. As suggested the peak time hours are analyzed, thus the timeframe is ranging between 13:00 to 16:00 where the sun is the highest, below the 25th percentile of the production level. The key parameters (variables to be examined and constants) are shown in table \ref{tab:data_columns}. The analysis of the examined data are described in detail in table \ref{tab:zarcoDes}. The 25th percentile corresponds to a production level below 18.25.

\begin{table}[H]
\centering
\small
\begin{tabular}{lr}
\hline
\textbf{Statistic} & \textbf{Value} \\
\hline
Count & 3656.00 \\
Mean & 29.22 \\
Standard Deviation (std) &  12.84 \\
Minimum (min) & 0 \\
25th Percentile & 18.25 \\
Median (50\%) &   33.25 \\
75th Percentile &  40.25 \\
Maximum (max) &   48.50 \\
\hline
\end{tabular}
\caption{Descriptive Statistics for production - Zarco Use case 1}
\label{tab:zarcoDes}
\end{table}
\subsubsection{EVA}

We filtered the data to include only these time-frames. EVA was then applied using the Peaks Over Threshold method, focusing on data below our defined threshold. This threshold was set to capture the lowest 25\% of production values, determined after statistical analysis of the data. To analyze the extreme value distribution, we fitted the data to a Generalized Pareto Distribution (GPD) model.

The histogram bars are not visible above the x-axis, which suggests that the observed values are so well-matched to the predicted values that the bars are hidden behind the PDF line. This would indicate an excellent fit if the theoretical model's PDF line accurately represents the observed histogram.

\begin{figure}
    \centering
    \includegraphics[width=0.9\linewidth]{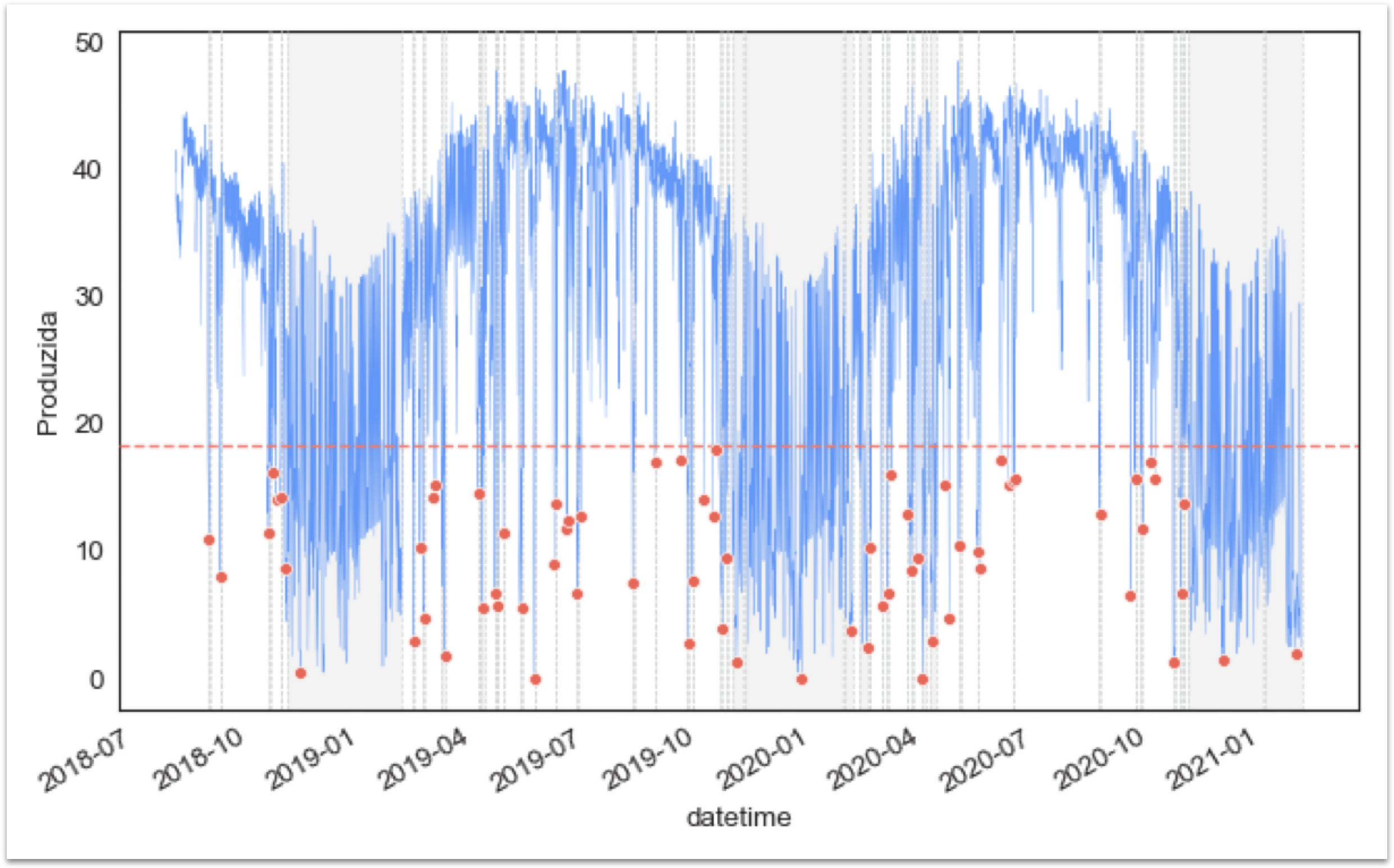}
    \caption{Peaks Under Threshold for "zarco" Use Case 1 }
    \label{fig:zarco}
\end{figure}

Return Periods (in years) can be seen in table \ref{tab:return_values_zarco}.

(1, 2, 5, 10, 25, ..) These numbers represent how often an event of a certain magnitude is expected to occur. For example, a 100-year return period event is something that, on average, we would expect to happen once every 100 years.

The Return Values represent the estimated magnitude of the event (e.g., production level) for each return period. The fact that these values are mostly negative suggests we're dealing with a scenario where the focus is on low production levels or deficits. As the return period increases, the return values tend to become more negative, indicating that more extreme deficits are less frequent.

Lower and Upper Confidence Intervals (CI): These values provide a range around the return value, within which the true value is expected to lie with a certain level of confidence. The confidence intervals get wider as the return period increases, indicating more uncertainty in predicting more extreme events.

1 year return period: The return value is 0.35 with a CI of [ 1.66, 0.11]. This means that in any given year, we can expect an extreme-low production level of around 0.35 during Peak hours, with a reasonable range of uncertainty between 1.66 and 0.11, and based on our analysis approximating 0 after 5 years.

\begin{table}[h!]
\centering 
\small
\begin{tabular}{|c|c|c|c|}
\hline
\textbf{Return Period} & \textbf{Return Value} & \textbf{Lower CI} & \textbf{Upper CI} \\
\hline
1.00  & 0.35 & 1.66  & 0.11  \\
2.00  & 0.15 & 1.31  & 0.04  \\
5.00  & 0.05 & 0.75  & 0.00  \\
10.00 & 0.02 & 0.58  & -0.20 \\
25.00 & 0.01 & 0.52  & -0.36 \\
\hline
\end{tabular}
\label{tab:return_values_zarco}
\caption{Return Periods and Confidence Intervals - Zarco Use Case 1}
\end{table}

\begin{figure}[H]
    \centering
    \includegraphics[width=0.9\linewidth]{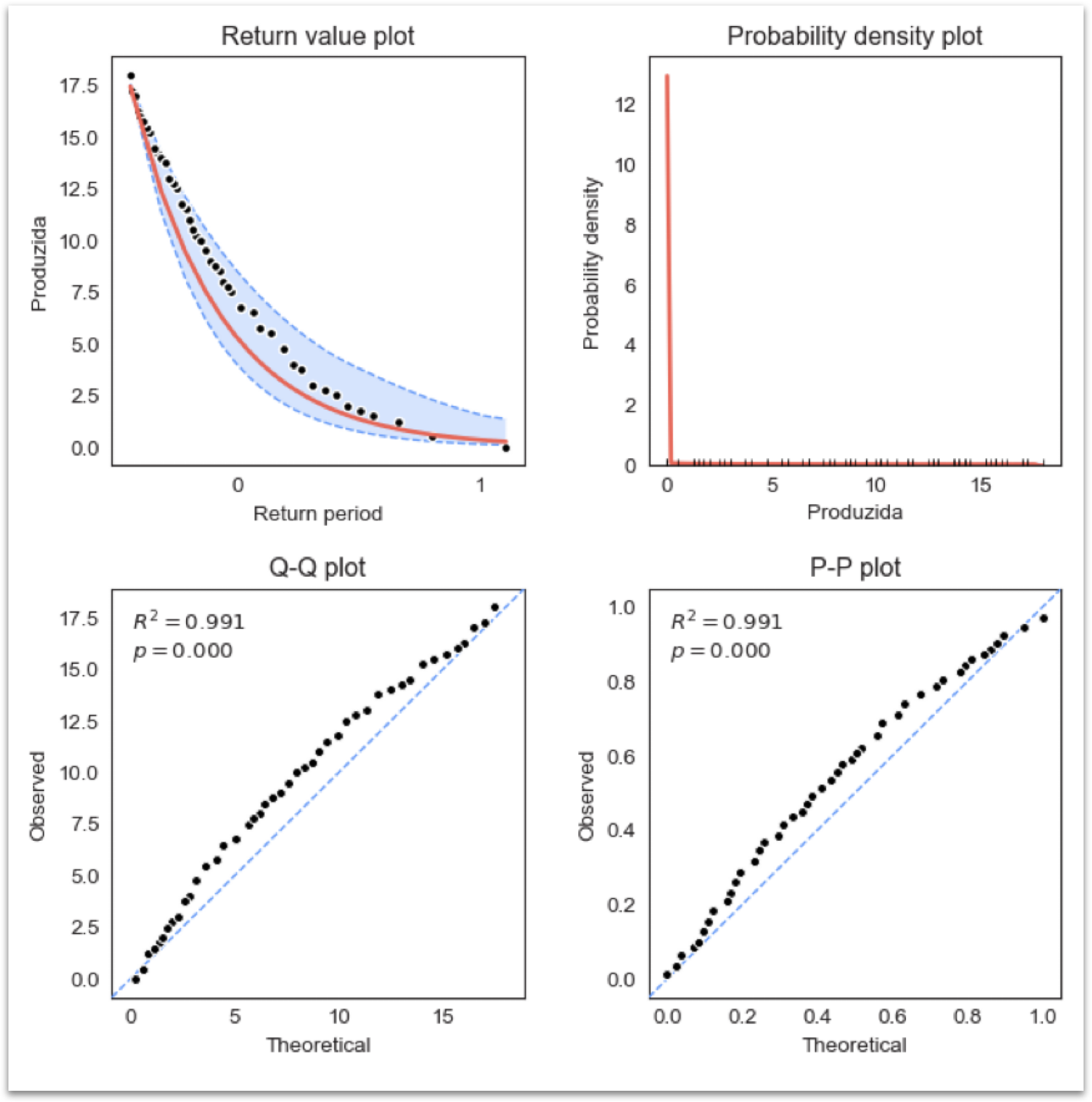}
    \caption{Return Plots and distributions for ``zarco"-Use Case 1}
    \label{fig:retursZarco}
\end{figure}

As can be seen in the return value plot \ref{fig:retursZarco} the data are well fitted within the distribution, thus the model is more reliable in predicting low production levels, allowing for more confidence on model’s predictive capabilities. 

The $R^2$ value of 0.997 and p-value of 0.000  confirm the model’s reliability in predicting the cumulative probabilities of low production events, and as shown in the plots shown in \ref{fig:retursZarco} suggesting a good fit .

\subsubsection{Circumstance Analysis}
The differences between the two scenarios are shown in table \ref{tab:joaostats}, when deducting the extremes

\begin{table}[h!]
\centering \small
\begin{tabular}{|l|c|c|c|c|c|c|c|}
\hline
\textbf{[table-\ref{tab:data_columns}]} & \textbf{v1} & \textbf{v2} & \textbf{v3} & \textbf{v4} & \textbf{v5} & \textbf{v6} & \textbf{v7}\\
\hline
\textbf{Count} & -3586.00 & -3586.00 & -3586.00 & -3586.00 & -3586.00 & -3586.00 & -3586.00 \\
\textbf{Mean} & 10.02 & -1.55 & 27.84 & 0.78 & -204.85 & 73.64 & -19.82 \\
\textbf{Min} & 23.00 & 5.00 & 0.00 & 3.00 & 32.30 & 20.64 & 0.00 \\
\textbf{25\%} & 9.00 & 1.00 & 22.00 & 1.00 & -147.80 & 63.51 & -12.75 \\
\textbf{50\%} & 6.00 & -1.00 & 50.50 & 1.00 & -237.93 & 113.13 & -23.50 \\
\textbf{75\%} & 14.00 & -3.00 & 38.75 & 0.00 & -339.21 & 104.32 & -26.25 \\
\textbf{Max} & -4.00 & -15.00 & 0.00 & -15.00 & -106.84 & -11.46 & -30.50 \\
\textbf{Std} & -0.81 & -2.18 & 1.71 & -0.52 & -57.07 & 8.25 & -7.55 \\
\hline
\end{tabular}
\caption{Statistical Summary of Weather and Production Data ``zarco" Use Case 1}
\end{table}

Applying the DISC-thresholding we extract the significant differences in correlations between the extreme low production and the normal production in the analyzed sample of peak time ranges (table \ref{tab:data_columns}). As can be seen in table  \ref{fig:corrJoao} significant differences are identified, for example between between Humidity and Diffuse Solar $w/m^2$ (HNN), humidity and temperature (HPC), for $\Delta \rho_{ij} = \rho_{ij}^{\text{extreme}} - \rho_{ij}$ and so on.

Following, the relevant information to the extremes are extracted to be used for benchmarking and comparison with a second Use Case to be analyzed in the next section. 

\subsection{EVDBM application Use Case 2: ``joao" Data}

\subsubsection{Data Synthesis}
\textbf{Report:} From a total of 21908 data, 3656 were related to peak time production 
During the peak hours the ours the data to be analyzed are 3652 with a mean production of 23.5 kwh. As suggested the peak time hours are analyzed, thus the timeframe is ranging between 13:00 to 16:00 where the sun is the highest, below the 25th percentile of the production level. The key parameters (variables to be examined and constants) are shown in table \ref{tab:data_columns}. The analysis of the examined data are described in detail in table \ref{tab:zarcoDes}. The 25th percentile corresponds to a production level below 13.43.

\begin{table}[H]
\centering \small
\begin{tabular}{lr}
\hline
\textbf{Statistic} & \textbf{Value} \\
\hline
Count & 3652.00 \\
Mean &   23.80 \\
Standard Deviation (std) &    11.51 \\
Minimum (min) &  0.25 \\
25th Percentile & 13.43 \\
Median (50\%) &   28.75 \\
75th Percentile &  33.75 \\
Maximum (max) &   40.00 \\
\hline
\end{tabular}
\caption{Descriptive Statistics for production - ``joao" Use case 2 `}
\label{tab:joaoDes}
\end{table}
\subsubsection{EVA}

Following the previous approach we filter the data as per the time range under the chosen percentile and apply EVA. 

\begin{figure}[H]
    \centering
    \includegraphics[width=0.9\linewidth]{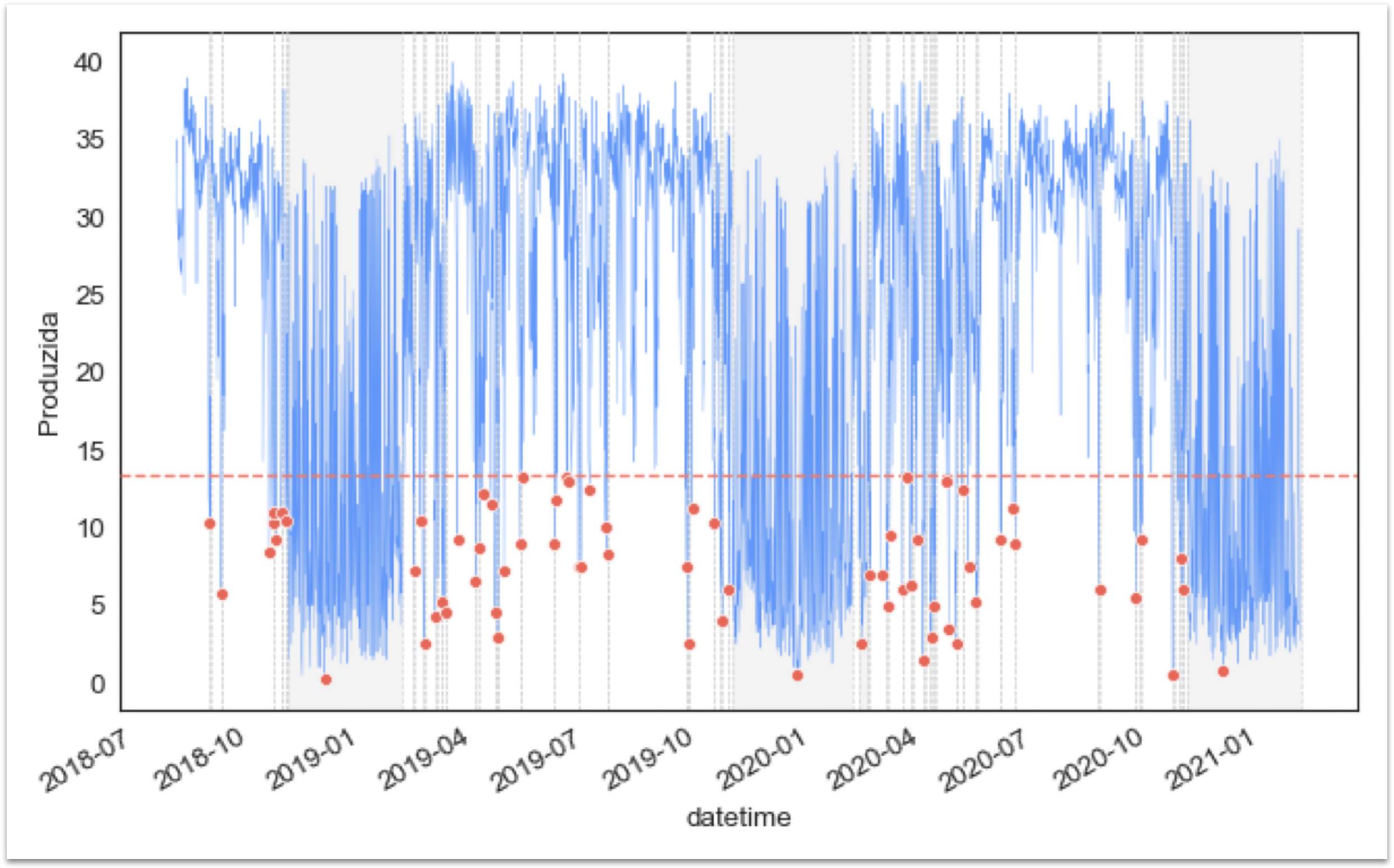}
    \caption{Peaks Under Threshold for ``joao" Use Case 2 }
    \label{fig:returnsjoao}
\end{figure}

As can be seen in the return value plot \ref{fig:returnsjoao} the data are well fitted within the distribution, thus the model is more reliable in predicting low production levels, allowing for more confidence on model’s predictive capabilities. 

Return Periods (in years) can be seen in table \ref{tab:return_values_joao}.
1 year return period: The return value is 1.05 with a CI of [ 1.96, 0.49]. This means that in any given year, we can expect an extreme-low production level of around 1.05 during Peak hours, with a reasonable range of uncertainty, significantly higher than the observed production of Use Case 1.

\begin{table}[h!]
\centering\small
\begin{tabular}{|c|c|c|c|}
\hline
\textbf{Return Period} & \textbf{Return Value} & \textbf{Lower CI} & \textbf{Upper CI} \\
\hline
1.00  & 1.05 & 1.96 & 0.49 \\
2.00  & 0.66 & 1.34 & 0.35 \\
5.00  & 0.39 & 0.91 & 0.18 \\
10.00 & 0.28 & 0.81 & -0.06 \\
25.00 & 0.20 & 0.72 & -0.23 \\
\hline
\end{tabular}
\label{tab:return_values_joao}
\caption{Return Periods and Confidence Intervals - Joao Use Case 2}
\end{table}

\begin{figure}[H]
    \centering
    \includegraphics[width=1\linewidth]{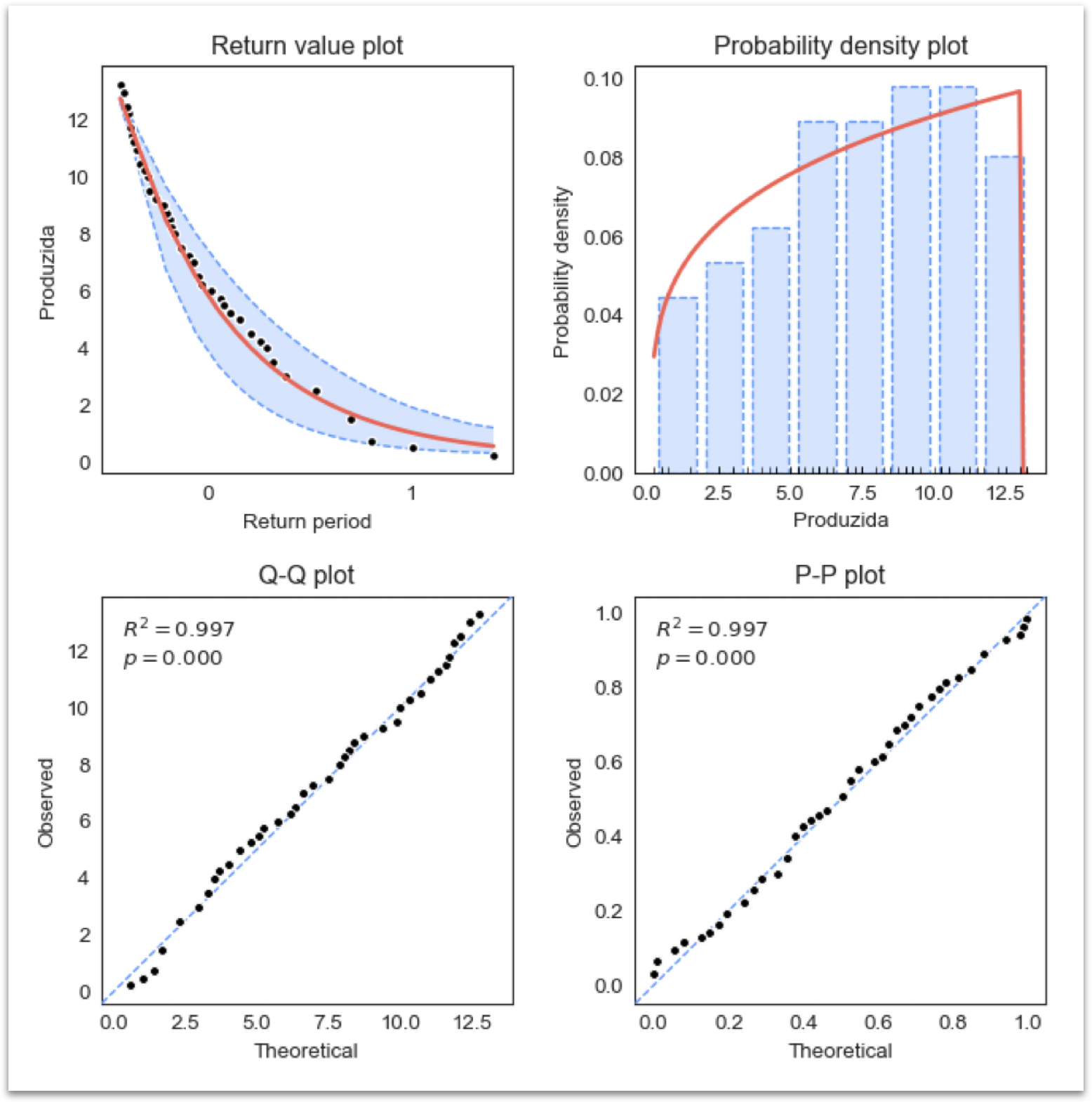}
    \caption{Return Plots and distributions for Joao - Use Case 2}
    \label{fig:returnsJoao}
\end{figure}

\subsubsection{Circumstance Analysis}
The differences between the two scenarios are shown in table \ref{tab:joaostats}

\begin{table}[h!]
\centering
\small
\begin{tabular}{|l|c|c|c|c|c|c|c|}
\hline
\textbf{[table-\ref{tab:data_columns}]} & \textbf{v1} & \textbf{v2} & \textbf{v3} & \textbf{v4} & \textbf{v5} & \textbf{v6} & \textbf{v7}\\
\hline
\textbf{Count} & -3583.00 & -3583.00 & -3583.00 & -3583.00 & -3583.00 & -3583.00 & -3583.00 \\
\textbf{Mean} & 11.32 & -1.81 & 30.48 & 0.48 & -194.70 & 50.41 & -16.33 \\
\textbf{Min} & 23.00 & 5.00 & 0.00 & 4.00 & 1.19 & 3.59 & 0.00 \\
\textbf{25\%} & 11.00 & 0.00 & 27.00 & 1.00 & -100.30 & 52.25 & -8.44 \\
\textbf{50\%} & 9.00 & -1.00 & 57.00 & 1.00 & -239.32 & 79.61 & -21.25 \\
\textbf{75\%} & 12.00 & -3.00 & 38.00 & -1.00 & -342.23 & 65.87 & -23.50 \\
\textbf{Max} & -1.00 & -16.00 & 0.00 & -3.00 & -86.26 & -26.15 & -26.75 \\
\textbf{Std} & -1.83 & -2.28 & 0.04 & 0.35 & -52.86 & 0.52 & -7.94 \\
\hline
\end{tabular}
\label{tab:joaostats}
\caption{Statistical Summary of Weather and Production Data - Use Case 2}
\end{table}

\begin{figure}[h!]
    \centering
    \includegraphics[width=1.1\linewidth]{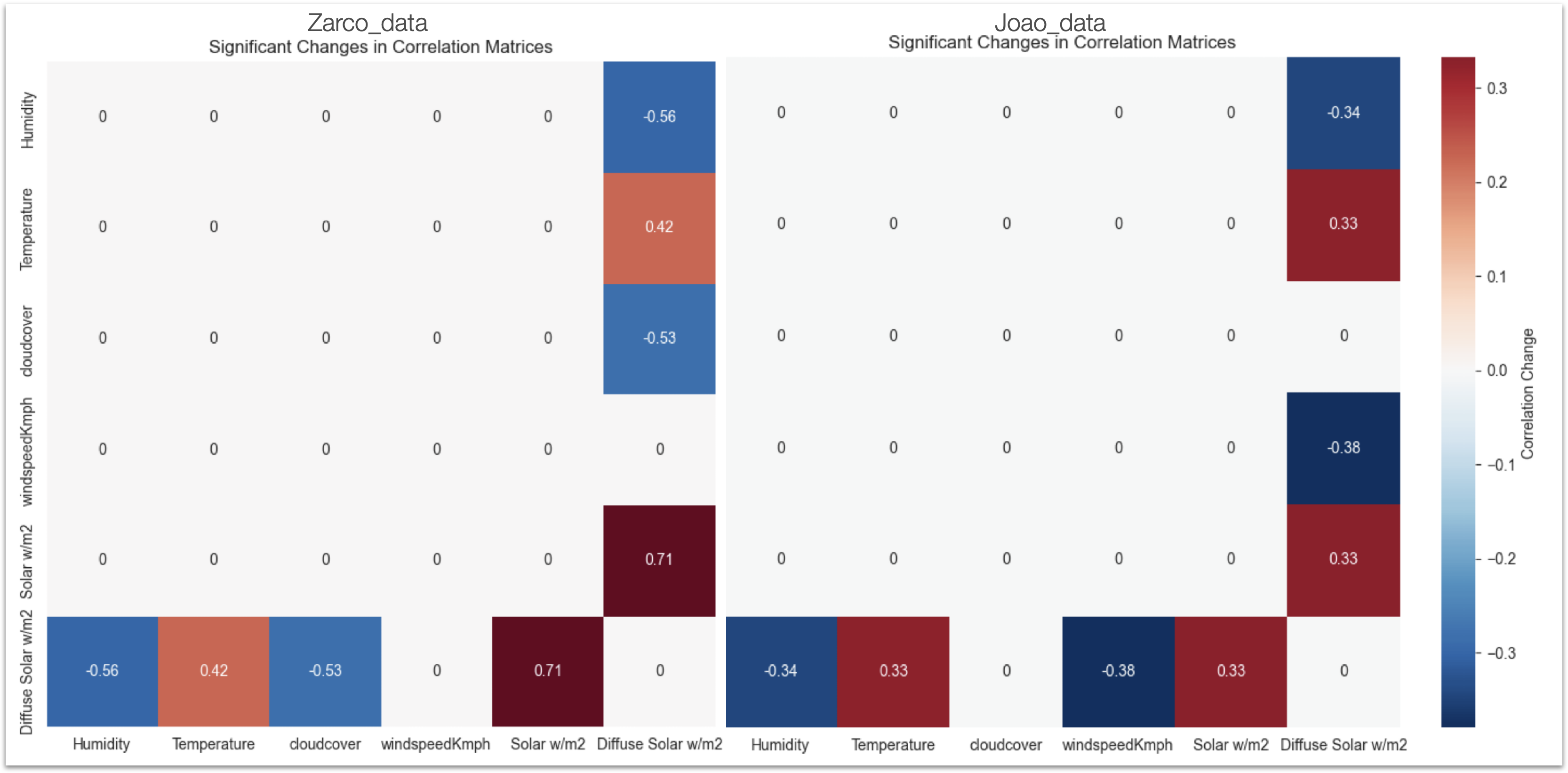}
    \caption{Significant Correlation differences between extreme and normal production levels for Peak times for Zarco and Joao Use Case 1 and 2 using the DISC Thresholding algorithm with a 90\% threshold}
    \label{fig:corrJoao}
\end{figure}

Applying the DISC-thresholding we extract the significant differences in correlations between the extreme low production and the normal production in the analyzed sample of peak time ranges (table \ref{tab:data_columns}). As can be seen in table  \ref{fig:corrJoao} significant differences are identified, for example between Humidity and Diffuse Solar $w/m^2$ (HNN), temperature and diffuse solar (HPC), for $\Delta \rho_{ij} = \rho_{ij}^{\text{extreme}} - \rho_{ij}$. (1)

\subsection{Benchmarking}
\begin{figure}[H]
    \centering
    \includegraphics[width=1.1\linewidth]{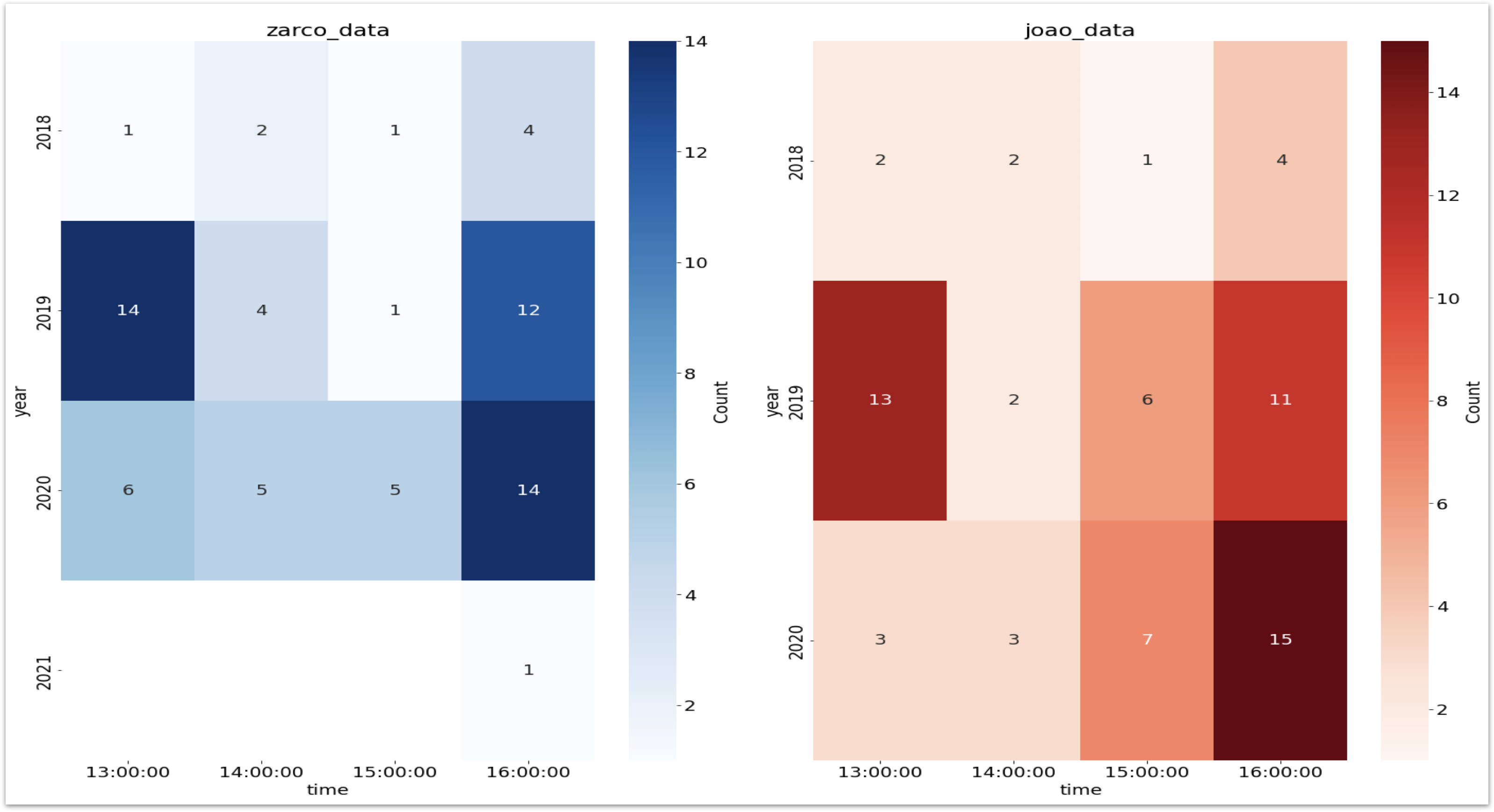}
    \caption{Extreme events grouped by year and time per Case}
    \label{fig:comparisonOB}
\end{figure}

In this section we extract the benchmarking score applying the methodology described in the previous section and plot the results to highlight the differences between the two examined scenarios. Figure \ref{fig:comparisonOB}
shows that for about similar cases, distribution per year and time is about the same. However using the EVA-Driven Weighted Benchmarking Algorithm. 

\begin{table}[h!]
\centering
\begin{tabular}{|l|c|c|}
\hline
\textbf{Variable} & \textbf{Suggested Weight $w_i$}  \\
\hline
Humidity (v1)& $w_1 = 0.05$  \\
Temperature (v2) & $w_2 = 0.1$ \\
Cloud Cover (v3) & $w_3 = 0.2$ \\
Windspeed (v4)& $w_4 = 0.05$ \\
Solar (Radiation) (v5)  & $w_5 = 0.4$ \\
Diffuse Solar (v6) & $w_6 = 0.2$  \\
\hline
\end{tabular}
\label{tab:weights}
\caption{Suggested Weights for Each Variable}
\end{table}

\begin{figure}
    \centering
    \includegraphics[width=1\linewidth]{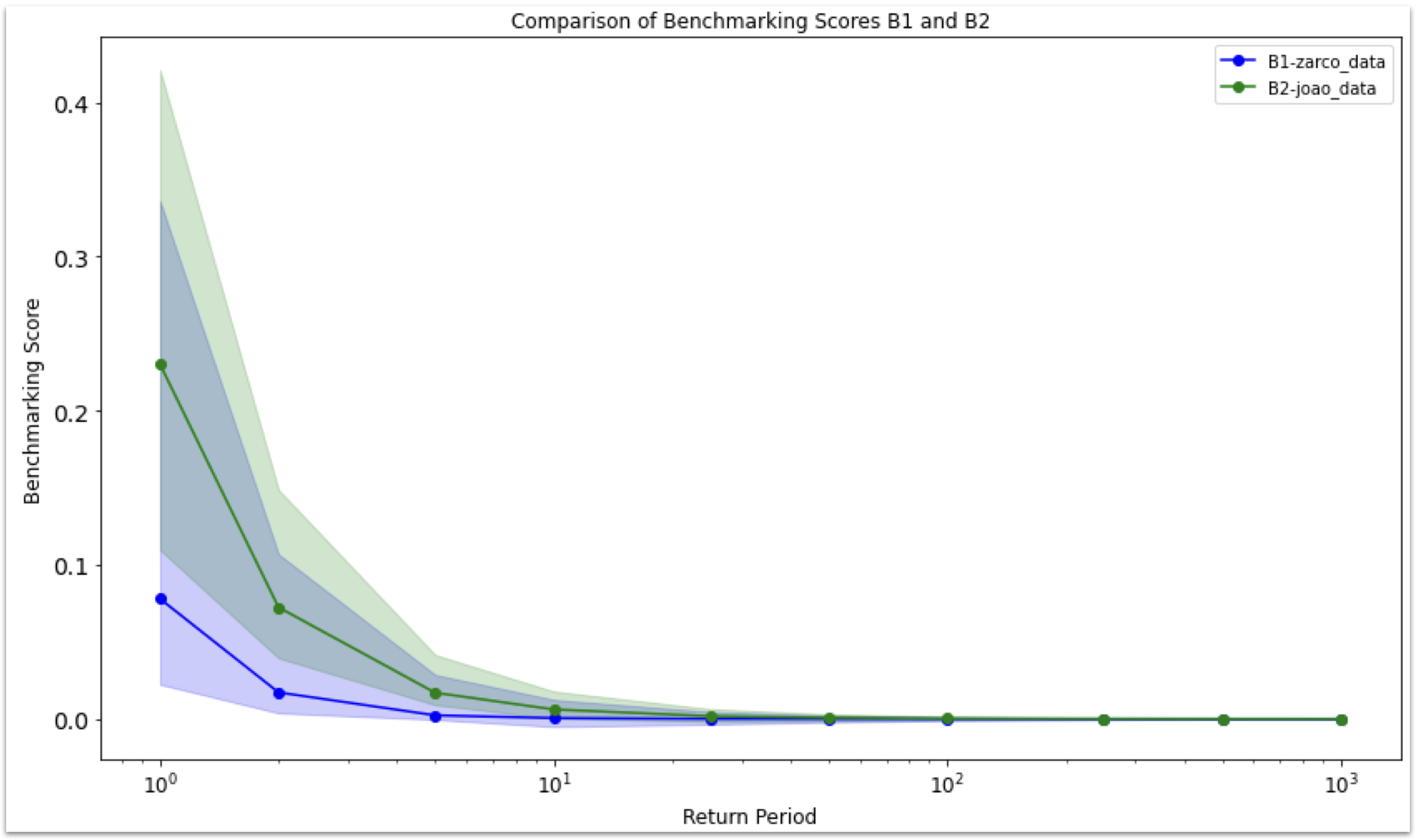}
    \caption{Extreme Value Dynamic Benchmarking Method (EVDBM) applied on photovoltaic energy production for analysis of related circumstances}
    \label{fig:plotsBench}
\end{figure}

Considering how each variable may impact the production for testing purposes we have defined some weights, presented in table \ref{tab:weights}. Solar radiation is given the highest weight because of its dominant influence on PV performance. Cloud cover and diffuse solar radiation are also heavily weighted, reflecting their significant roles, especially during low production periods. Temperature, windspeed, and humidity have lower weights, as their impacts are secondary but still important to consider. 
The normalized, weighted and scaled results following the methods described are plotted in figure \ref{fig:plotsBench}

The benchmarking score reflects how strongly the conditions (like humidity, sunlight, wind speed) contribute to low production events (extreme low production values identified by EVA), as low production was the basis of this analysis. A \textit{high benchmarking} score suggests that the combination of environmental conditions has a greater impact on driving the system toward extreme low production. In other words, when these conditions align in specific ways, they correlate strongly with low production events. A high score implies that the system (PV plant) is more sensitive or vulnerable to these environmental factors under extreme conditions. The higher score indicates that extreme low production is likely to occur when these specific conditions are met, showing a greater dependency on or sensitivity to adverse conditions. Thus, a higher score  indicates less resilience to environmental variations. 

Since B2 has the highest EVDBM score it is clear that B2 (joao plant) is more sensitive to adverse conditions and is less resilient to fluctuations in related circumstances. 


\section{Discussion of Results}

In this paper, we expanded Extreme Value Analysis (EVA) by incorporating additional tools to create a more streamlined and adaptable analytical approach. We introduced the Extreme Value Dynamic Benchmarking Method (EVDBM), a novel framework for analyzing extreme events across multiple domains. By leveraging both the Peaks Over Threshold (POT) and Block Maxima (Minima) methods, this approach enables flexible event detection. In addition, the integration of the Dynamic Identification of Significant Correlation (DISC)-Thresholding algorithm provides a more refined analysis of how key variables behave under extreme conditions. The EVA-Driven Weighted Benchmarking Algorithm further enhances performance comparison by weighting variables based on their influence during extreme events. This comprehensive framework deepens the understanding of extreme occurrences and their impacts on system performance, while also enabling continuous monitoring through extracted overall scores. By tracking these scores over time, the methodology allows for the ongoing evaluation of extreme events, informing decision-making and facilitating adaptation as conditions change.

By integrating return values predicted through Extreme Value Analysis (EVA) into the benchmarking scores, we are able to transform these scores to reflect anticipated conditions more accurately. This approach provides a more precise picture of how each case is projected to unfold under extreme conditions. The use of EVA-based return values allows us to model the expected \textit{intensity} and \textit{likelihood} of extreme events over different timeframes, which in turn reveals how each related circumstance is likely to develop. As a result, the adjusted scores offer a forward-looking perspective, highlighting potential vulnerabilities and resilience factors for each case in a way that static historical data alone cannot capture.

In the photovoltaic (PV) energy use case, the method successfully captured critical low-production events and identified significant correlations between variables, illustrating its practical application for managing operational risks in renewable energy. Furthermore, it enabled the comparison of different use cases through the generation of an overall score. This flexibility highlights the potential of the methodology for broader applications in other fields where extreme value assessment and benchmarking under uncertain conditions are required.

In the case of PV energy production for example this approach could aid in decision making in a variety of ways like:
\begin{itemize}
\item \textbf{Predicting Low Energy Events:} EVA estimates energy production for various return periods, identifying rare, low-output events for photovoltaic plants \cite{hegerl2018early,shukla1998predictability}.
\item \textbf{Risk Management:} Knowledge of low-production frequencies aids in planning and mitigating risks through diversification, storage, and demand response \cite{hilorme2019formation,mishra2016bridging,talluri2013assessing}.

\item \textbf{Infrastructure and Investment:} Low-output scenarios inform PV design, backup needs, and financial planning \cite{truong2016s,elmassri2016accounting,harrison1977decision,guthrie2019real}.

\item \textbf{Confidence Intervals:} Confidence intervals highlight uncertainty ranges, aiding scenario planning for risk management \cite{pagano2018dealing,lind2002fairness}.

\item \textbf{Climate Change Adaptation:} EVA helps in assessing impacts of climate patterns on solar production, supporting adaptation strategies \cite{van2021communicated,thaler2021governance,moser2010framework,biesbroek2013nature}.

\item \textbf{Policy and Compliance:} EVA data informs policies on reserve capacity and renewable credits.

\item \textbf{Benchmarking PV Plants:} EVDBM scores and compares PV plants based on climate factors, aiding in performance optimization and investment decisions.
\end{itemize}
\subsection{Limitations and future work}
The main limitations of this work to be addressed are the following:

\begin{itemize}
    \item Sensitivity to Data Quality and Availability: EVDBM relies heavily on robust historical data, particularly for extreme events, to make accurate predictions. In cases where data on past extremes is limited or unavailable, predictions and benchmarking scores may lack accuracy
    \item Assumption of Stationarity partially addressed: External changes could alter the probability and severity of future extremes, impacting the reliability of benchmarks based on historical data.
    \item Subjectivity in Weighting Variables: If weighting isn’t carefully calibrated, it can introduce bias, making some cases appear more or less resilient than they might be in practice. This can particularly impact comparisons across cases with differing sensitivity to certain variables
\end{itemize}

In our future work we intend to apply this methodology in other use cases where increased specificity in decision making can lead to optimized and beneficial outcomes, such as in health care and in finance. We also work on integrating EVDBM algorithm for scenario-based analysis, what-if scenarios and other risk-management applications while attaching a more dynamic approach to related circumstances prediction.
\newpage

\section*{Acknowledgments}
The work presented is based on research conducted within the framework of the Horizon Europe European Commission project CRETE VALLEY (Grant Agreement No. 101136139) and the Horizon Europe European Commission project “Energy Activated Citizens and Data-Driven Energy-Secure Communities for a Consumer-Centric Energy System (ENPOWER)” under grant agreement no. 101096354. The content of the paper is the sole responsibility of its authors and does not necessarily reflect the views of the EC.

\section*{Appendix}
The following abbreviations are used in this paper:\\
\begin{table}[H]
    \centering
    \begin{tabular}{cl}
       EVA & Extreme Value Analysis \\
        EVT & Extreme Value Theory \\
        PV & PhotoVoltaic \\
       EVDBM  & Extreme Value Dynamic Benchmarking Method\\
       DISC & Dynamic Identification of Significant Correlation \\
       POT & peak-over-threshold \\
       R.V & return level \\
       PDF & Probability Density Function \\
        HPC & High Positive Correlation \\
        HNN & and a High Negative Correlation \\
    \end{tabular}
    \label{tab:my_label}
\end{table}

 \bibliographystyle{elsarticle-num} 

\bibliography{ref}

\end{document}